 \documentclass[traditabstract]{aa} % for the abstract without structuration
\usepackage{graphicx}
\usepackage{txfonts}
\usepackage{natbib}
\usepackage{amsmath}
\usepackage{ulem} 
\usepackage{lscape}

\usepackage{indentfirst}
\bibpunct[]{(}{)}{;}{a}{}{,}

\usepackage{threeparttablex} % for "ThreePartTable" environment
\usepackage{booktabs}        % for well-spaced horizontal rules

\begin{document}  

   \title{Rotational spectroscopy of methyl mercaptan CH$_3$$^{32}$SH at millimeter and submillimeter
          wavelengths\thanks{This manuscript is dedicated to the memory of Li-Hong Xu who passed away 
          at the final stage of writing of the manuscript.}$^,$\thanks{The line files of the MW data 
          and of the FTFIR data along with a prediction file up to 2~THz are available as text files 
          at CDS via anonymous ftp to cdsarc.u-strasbg.fr 
         (130.79.128.5) or via http://cdsweb.u-strasbg.fr/cgi-bin/qcat?J/A+A/.}}

   \author{Olena Zakharenko\inst{1}
           \and
           Vadim V. Ilyushin\inst{2,3}
           \and
           Frank Lewen\inst{1}
           \and
           Holger S.~P. M{\"u}ller\inst{1}
           \and
           Stephan Schlemmer\inst{1}
           \and
           Eugene A. Alekseev\inst{2,3}
           \and
           Mykola L. Pogrebnyak\inst{2,3}
           \and
           Iuliia A. Armieieva\inst{2}
           \and
           Olha Dorovskaya\inst{2}
           \and
           Li-Hong Xu\inst{4}
           \and
           Ronald M. Lees\inst{4}
           }

   \institute{I.~Physikalisches Institut, Universit{\"a}t zu K{\"o}ln,
              Z{\"u}lpicher Str. 77, 50937 K{\"o}ln, Germany\\
              \email{zakharenko@ph1.uni-koeln.de, hspm@ph1.uni-koeln.de}
              \and
              Institute of Radio Astronomy of NASU, Mystetstv 4, 61002 Kharkiv, Ukraine\\
              \email{ilyushin@rian.kharkov.ua}
              \and
              Quantum Radiophysics Department, V.~N. Karazin Kharkiv National University, Svobody Square 4, 61022 Kharkiv, Ukraine
              \and
              Department of Physics, University of New Brunswick, Saint John, NB, Canada
              }

   \date{Received 24 April 2019 / Accepted 25 July 2019}

%%%%%%%%%%%%%%%%%%%%%%%%%%%%%%%%%%%%%%%%%%%%%%%%%%%%%%%%%%%%%%%%%%%%%%%%%%%%%%%%%%%%%%%%%
%%%%%%%%%%%%%%%%%%%%%%%%%%%%%%%%%%%%%%%%%%%%%%%%%%%%%%%%%%%%%%%%%%%%%%%%%%%%%%%%%%%%%%%%%
  \abstract
%%%%%%%%%%%%%%%%%%%%%%%%%%%%%%%%%%%%%%%%%%%%%%%%%%%%%%%%%%%%%%%%%%%%%%%%%%%%%%%%%%%%%%%%%
%%%%%%%%%%%%%%%%%%%%%%%%%%%%%%%%%%%%%%%%%%%%%%%%%%%%%%%%%%%%%%%%%%%%%%%%%%%%%%%%%%%%%%%%%

\abstract{We present a new global study of the millimeter (mm) wave, submillimeter (sub-mm) wave, 
and terahertz (THz) spectra of the lowest three torsional states of methyl mercaptan (CH$_3$SH). 
New measurements have been carried out between 50 and 510~GHz using the Kharkiv mm wave and 
the Cologne sub-mm wave spectrometers whereas THz spectra records were used from our previous study. 
The new data, involving torsion-rotation transitions with $J$ up to 61 and $K_a$ up to 18, 
were combined with previously published measurements and fit using the rho-axis-method 
torsion-rotation Hamiltonian. The final fit used 124 parameters to give an overall weighted 
root-mean-square deviation of 0.72 for the dataset consisting of 6965 microwave (MW) and 
16345 far-infrared line frequencies sampling transitions within and between the ground, first, 
and second excited torsional states. This investigation presents a two-fold expansion in the 
$J$ quantum numbers and a significant improvement in the fit quality, especially for the MW 
part of the data, thus allowing us to provide more reliable predictions to support 
astronomical observations.}


\authorrunning{Olena Zakharenko et al.}
\titlerunning{Laboratory spectroscopy of CH$_3$$^{32}$SH}

\maketitle
\hyphenation{For-schungs-ge-mein-schaft}

%%%%%%%%%%%%%%%%%%%%%%%%%%%%%%%%%%%%%%%%%%%%%%%%%%%%%%%%%%%%%%%%%%%%%%%%%%%%%%%%%%%%%%%%%
%%%%%%%%%%%%%%%%%%%%%%%%%%%%%%%%%%%%%%%%%%%%%%%%%%%%%%%%%%%%%%%%%%%%%%%%%%%%%%%%%%%%%%%%%
\section{Introduction}
\label{intro}
%%%%%%%%%%%%%%%%%%%%%%%%%%%%%%%%%%%%%%%%%%%%%%%%%%%%%%%%%%%%%%%%%%%%%%%%%%%%%%%%%%%%%%%%%
%%%%%%%%%%%%%%%%%%%%%%%%%%%%%%%%%%%%%%%%%%%%%%%%%%%%%%%%%%%%%%%%%%%%%%%%%%%%%%%%%%%%%%%%%

Sulfur-bearing interstellar molecules are of interest for astrophysics since their abundance 
is particularly sensitive to the physical and chemical evolution in the warm and dense parts 
of star-forming regions, called hot cores or hot corinos. Their molecular ratios are used 
as chemical clocks to obtain information about the age of these regions  
\citep{Charnley1997,Hatchell1998,1998Hatchell,Wakelam2011}. At the same time, the systematic 
understanding of interstellar sulfur chemistry is not yet achieved because of the so-called 
sulfur depletion problem \citep{Ruffle1999}. Much less sulfur is found in dense regions 
of the interstellar medium than in diffuse regions \citep{Anderson:2013}, and there is 
some problem concerning this missing sulfur and what might be its reservoir. 
Therefore, extension of observations for interstellar sulfur bearing molecules are of interest 
for a better understanding of the star-formation process.

Methyl mercaptan (CH$_3$SH), also known as methanethiol, is an important sulfur-bearing species 
not only in the interstellar medium, but also for the terrestrial environment, and potentially 
in planetary atmospheres \citep{Vance:2011}. 
It was first tentatively detected in Sgr B2 by \citet{Turner:1975} and then definitively confirmed 
by \citet{Linke139L}. Later, methyl mercaptan was observed toward the high-mass star-forming
region G327.3$-$0.6 \citep{Gibb:2000}, the cold core B1 \citep{2012ApJ...759L..43C}, the Orion~KL 
hot core \citep{2014ApJ...784L...7K}, the low-mass star-forming region IRAS 16293$-$2422 
\citep{2016MNRAS.458.1859M}, and the prestellar core L1544 \citep{2018MNRAS.478.5514V}. 
Recently, methyl mercaptan was observed in molecular line surveys carried out with the 
Atacama Millimeter/submillimeter Array (ALMA) towards Sgr B2(N2) and IRAS 16293$-$2422 at levels 
that make detection of some of its isotopologs probable \citep{Holger2016,PILS_2018}. 
A recent search \citep{2019A&A...621A.114Z} for CH$_3$SD toward the solar-type proto-star 
IRAS 16293$-$2422 B, however, was negative even though the upper limit to the column density 
of CH$_3$SD may have been close to the expected value.

The main isotopic species CH$_3$$^{32}$SH was subjected to numerous spectroscopic studies 
mainly from the perspectives of torsional large amplitude motion investigations. 
The rich and complex torsion-rotation dynamics, characterized by a relatively large coupling 
term between internal rotation and global rotation in this molecule, provides a good test case 
for different theoretical models in use. Early investigations of the methyl mercaptan rotational 
spectrum were carried out about 60 years ago \citep{Solimene:1955,Kojima:1957,Kojima:1960}. 
These investigations were extended later into the millimeter (mm) and lower submillimeter (sub-mm) 
wave regions \citep{Lees:1980,Sastry:1986,Bettens:1999}. 
The most recent works extended investigations further into the terahertz (1.1$-$1.8~THz) and 
far-infrared (FIR) regions (50$-$560~cm$^{-1}$) \citep{Xu:2012, 2018JMoSp.352...45}.

Despite the fact that significant progress was achieved in understanding the rotational spectra of 
the lowest three torsional states of methyl mercaptan \citep{Xu:2012}, some problems in fitting 
the microwave (MW) data remained. Whereas the overall weighted root mean square (rms) deviation 
of the fit was 1.071, the weighted rms deviation of the MW data was 2.586, ranging from 2.075 
in the ground torsional state to 4.369 in the second excited torsional state \citep{Xu:2012}. 
We decided to address this problem by initiating a new global study of the mm wave, sub-mm wave, 
and THz spectra of the lowest three torsional states of methyl mercaptan. While we apply the same 
rho axis method (RAM) approach \citep{Kirtman:1962,Lees:1968,HOUGEN:1994}, the computer program 
used here is different from the previous RAM study of methyl mercaptan spectrum \citep{Xu:2012}, 
where a version of the BELGI code \citep{Kleiner:2010} was used that is described 
in some detail in \citet{XU:2008305}. In the current study, we employ the RAM36 
(rho-axis-method for 3- and 6-fold barriers) code \citep{Ilyushin:2010,Ilyushin:2013} 
that provides the opportunity to choose almost any symmetry-allowed term in the Hamiltonian 
and thus extends the RAM parameter space available for exploration in comparison with BELGI 
\citep{Kleiner:2010,XU:2008305}. This opportunity, as well as the proper treatment of blends 
built in the RAM36 program, were the main arguments in favor of moving to the RAM36 code 
platform in the current study of the methyl mercaptan spectrum. This treatment of blends 
was absent in the earlier version of the BELGI code \citep{XU:2008305} used previously 
to fit the methyl mercaptan spectrum \citep{Xu:2012}. Such blends are frequently caused 
by unresolved $K$-doublets of A symmetry lines or by accidental overlap of transitions. 
They were expected to be part of the problem with the rather high weighted rms deviation 
of MW lines in \citet{Xu:2012}.  In addition, we decided to extend the $J$ quantum number 
coverage. With this aim, new measurements were carried out between 50 and 510~GHz 
in addition to the terahertz (1.1$-$1.8~THz) spectrum records, which were available for 
analysis from the previous study \citep{Xu:2012}. Our ultimate goal was to extend reliable 
predictions of the CH$_3$$^{32}$SH spectrum to support astronomical observations by radio 
telescopes in particular at mm and sub-mm wavelengths.

%%%%%%%%%%%%%%%%%%%%%%%%%%%%%%%%%%%%%%%%%%%%%%%%%%%%%%%%%%%%%%%%%%%%%%%%%%%%%%%%%%%%%%%%%
%%%%%%%%%%%%%%%%%%%%%%%%%%%%%%%%%%%%%%%%%%%%%%%%%%%%%%%%%%%%%%%%%%%%%%%%%%%%%%%%%%%%%%%%%
\section{Experimental details}
\label{exptl}
%%%%%%%%%%%%%%%%%%%%%%%%%%%%%%%%%%%%%%%%%%%%%%%%%%%%%%%%%%%%%%%%%%%%%%%%%%%%%%%%%%%%%%%%%
%%%%%%%%%%%%%%%%%%%%%%%%%%%%%%%%%%%%%%%%%%%%%%%%%%%%%%%%%%%%%%%%%%%%%%%%%%%%%%%%%%%%%%%%%

Measurements in Cologne were done in frequency ranges within 155$-$510 GHz using the Cologne 
mm/sub-mm wave spectrometer. An Agilent E8257D synthesizer, referenced to a rubidium standard, 
together with an appropriate VDI (Virginia Diodes, Inc.) amplified multiplier chain, 
were used as a frequency source. The output mm/sub-mm radiation was directed to the 5~m 
double-pass glass cell of 10~cm diameter and then to the detectors. We used Schottky diode 
detectors to detect the output signal. The measurements were carried out at room temperature 
and at pressures of 20$-$40~$\mu$bar. The input frequency was modulated at 47.8~kHz. 
The modulation amplitude and frequency steps were adjusted to optimize the signal-to-noise 
ratio (S/N). The output signal from the detectors was detected by a lock-in amplifier in 
$2f$ mode to give second-derivative spectra, with a time constant of 20 or 50~ms. 
A detailed description of the spectrometer may be found in \citet{Bossa:2014} and 
\citet{Xu:2012}. Methyl mercaptan ($\ge 98.0\%$) was purchased from Sigma Aldrich and 
used without further purification.

Measurements in Kharkiv were done in the frequency range of 49$-$150~GHz using the 
automated spectrometer of the Institute of Radio Astronomy of NASU \citep{Alekseev:2012}. 
The synthesis of the frequencies in the mm wave range is carried out by a two-step 
frequency multiplication of a reference synthesizer in two phase-lock-loop (PLL) stages. 
The reference synthesizer is a computer-controlled direct digital synthesizer (DDS AD9851), 
whose output is up-converted into the 385$-$430~MHz frequency range. A klystron operating 
in the 3.4$-$5.2~GHz frequency range with a narrowband (1~kHz) PLL system is used at the 
first multiplication stage. An Istok backward wave oscillator (BWO) is locked to a harmonic 
of the klystron at the second multiplication stage. A set of BWOs is used to cover the frequency 
range from 49 to 149 GHz. The input frequency was modulated at 11.16 kHz, and the output signal 
from the detectors was detected by a lock-in amplifier in $1f$ mode to give first derivative 
spectra. The measurements were carried out at room temperature and at pressures of 
10$-$20~$\mu$bar. The uncertainties of the measurements were estimated to be 10~kHz for 
a relatively strong isolated line (S/N > 10), 30~kHz for weak lines (2 < S/N < 10) 
and 100~kHz for very weak lines (S/N < 2). Methyl mercaptan was synthesized by adding HCl to 
a 21\% water solution of sodium thiomethoxide CH$_3$SNa (purchased from Sigma Aldrich and 
used without further purification). There was no clear indication of lines of side-products 
in the Kharhiv spectral recordings except few rather weak water lines at known positions.

%%%%%%%%%%%%%%%%%%%%%%%%%%%%%%%%%%%%%%%%%%%%%%%%%%%%%%%%%%%%%%%%%%%%%%%%%%%%%%%%%%%%%%%%%
%%%%%%%%%%%%%%%%%%%%%%%%%%%%%%%%%%%%%%%%%%%%%%%%%%%%%%%%%%%%%%%%%%%%%%%%%%%%%%%%%%%%%%%%%
\section{Theoretical model}
\label{model}
%%%%%%%%%%%%%%%%%%%%%%%%%%%%%%%%%%%%%%%%%%%%%%%%%%%%%%%%%%%%%%%%%%%%%%%%%%%%%%%%%%%%%%%%%
%%%%%%%%%%%%%%%%%%%%%%%%%%%%%%%%%%%%%%%%%%%%%%%%%%%%%%%%%%%%%%%%%%%%%%%%%%%%%%%%%%%%%%%%%

In the current study, we used the so-called rho-axis-method \citep{HOUGEN:1994}, which was 
already applied successfully to the analysis of the methyl mercaptan spectrum in the past 
\citep{Xu:2012}. The Hamiltonian is based on the work of \citet{Kirtman:1962}, 
\citet{Lees:1968}, and \citet{Herbst:1984} and proved its effectiveness for a number of 
molecules containing a $C_{3v}$ rotor and a $C_s$ frame. While we applied the same method, 
the computer program employed here is different from the previous RAM studies of 
the methyl mercaptan spectrum, where the BELGI code was used \citep{Kleiner:2010,XU:2008305}. 
We chose the RAM36 code \citep{Ilyushin:2010,Ilyushin:2013} for the present analysis 
of the spectra. It provides enhanced calculation performance in comparison with 
the BELGI code. 

The general expression of the RAM Hamiltonian, that allows a global fit of the ground 
torsional state together with excited torsional states, may be written as follows:

%\begin{equation}
\begin{gather*}
H=1/2\sum_{pqnkstl}B_{pqnkstl}\{J^{2p},J_z^q,J_x^n,J_y^k,p_{\alpha}^s,\cos(3t\alpha),\sin(3l\alpha)\},  \quad(1)  
\end{gather*}
%\end{equation}    

\noindent 
where the $B_{pqnkstl}$ are fitting parameters; $p_\alpha$ is the angular momentum conjugate 
to the internal rotation angle $\alpha$; $J_x$, $J_y$, $J_z$ are projections on the 
$x$, $y$, $z$ axes of the total angular momentum $J$, and \{A,B,C,D,E,F,G\} = 
ABCDEFG + GFEDCBA is a generalized anticommutator. The RAM36 code provides an opportunity 
to choose almost any symmetry-allowed term in the Hamiltonian (by choosing an appropriate set 
of $k$, $n$, $p$, $q$, $l$, $s$, $t$ integer indices in Eq. (1)). 
The RAM36 code uses the two step diagonalization procedure of \citet{Herbst:1984}, and 
in the current study, we keep 21 torsional basis functions at the first diagonalization 
step and 11 torsional basis functions at the second diagonalization step. A more detailed 
description of the RAM36 code can be found in \citet{Ilyushin:2010,Ilyushin:2013}.

%%%%%%%%%%%%%%%%%%%%%%%%%%%%%%%%%%%%%%%%%%%%%%%%%%%%%%%%%%%%%%%%%%%%%%%%%%%%%%%%%%%%%%%%%
%%%%%%%%%%%%%%%%%%%%%%%%%%%%%%%%%%%%%%%%%%%%%%%%%%%%%%%%%%%%%%%%%%%%%%%%%%%%%%%%%%%%%%%%%
    
\begin{figure*}[t]
 \centering
 \includegraphics[height=10cm]{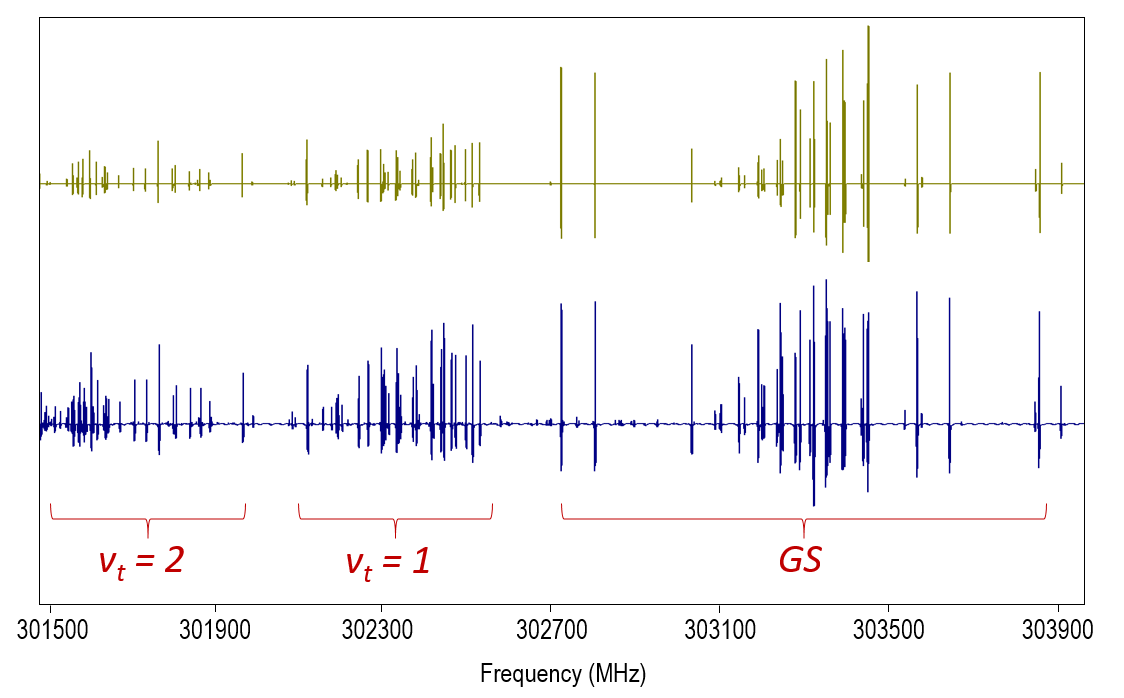}
 \caption{Detail of the CH$_3$SH rotational spectrum showing $a$-type $R$-branch transitions 
   with $J =12 \leftarrow 11$. The experimental spectrum is shown in the lower trace. 
   A simulation of CH$_3$SH transitions up to $\varv_t = 2$ is shown in the upper trace.}
 \label{fgr:mmWaveSpektrum}
\end{figure*}

%%%%%%%%%%%%%%%%%%%%%%%%%%%%%%%%%%%%%%%%%%%%%%%%%%%%%%%%%%%%%%%%%%%%%%%%%%%%%%%%%%%%%%%%%
%%%%%%%%%%%%%%%%%%%%%%%%%%%%%%%%%%%%%%%%%%%%%%%%%%%%%%%%%%%%%%%%%%%%%%%%%%%%%%%%%%%%%%%%% 

It should be noted that we have modified the labeling scheme of the RAM36 code 
for the current study to be conform with the labeling scheme used for methyl mercaptan 
in the previous study \citep{Xu:2012}. The standard labeling scheme in RAM36, after 
the second diagonalization step, begins by using eigenfunction composition to determine 
the torsional state to which a particular level belongs, and then uses the usual 
asymmetric-rotor energy ordering scheme to assign rotational $K_a$, $K_c$ labels 
within a given torsional state. This provides a relatively robust and simple labeling 
scheme in a case when a dominant basis function in the eigenvector composition is absent 
due to extensive basis-set mixing. In the CH$_3$SH molecule, which is a nearly symmetric 
prolate top ($\kappa$ = $-$0.988), the angle between the rho-axis-method $a$-axis and the
principal-axis-method $a$-axis is only 0.14$^\circ$. Such a small angle means that the 
RAM $a$-axis in methyl mercaptan is quite suitable for $K$ quantization and that 
eigenvectors can be unambiguously assigned using dominant eigenvector component. 
Thus, we label levels based on eigenfunction composition in the current study by searching 
for a dominant eigenvector component as it was done in the study of \citet{Xu:2012}. 
The energy levels in this work are labeled by free rotor quantum number $m$, overall 
rotational angular momentum quantum number $J$, and a signed value of $K_a$, 
which is the axial $a$-component of the overall rotational angular momentum $J$. 
In the case of A species, the $+/-$ sign corresponds to the so-called "parity" 
designation, which is in a slightly complicated way related to the A1/A2 
symmetry species in $G_6$ \citep{HOUGEN:1994}. For the E species, 
the signed value of $K_a$ reflects the fact that the Coriolis-type interaction 
term between the internal rotation and the global rotation causes the $|K_a|$ 
levels to split into $K_a$ > 0 and $K_a$ < 0 levels.

%%%%%%%%%%%%%%%%%%%%%%%%%%%%%%%%%%%%%%%%%%%%%%%%%%%%%%%%%%%%%%%%%%%%%%%%%%%%%%%%%%%%%%%%%
%%%%%%%%%%%%%%%%%%%%%%%%%%%%%%%%%%%%%%%%%%%%%%%%%%%%%%%%%%%%%%%%%%%%%%%%%%%%%%%%%%%%%%%%%
\section{Assignments and fit}
\label{fit}
%%%%%%%%%%%%%%%%%%%%%%%%%%%%%%%%%%%%%%%%%%%%%%%%%%%%%%%%%%%%%%%%%%%%%%%%%%%%%%%%%%%%%%%%%
%%%%%%%%%%%%%%%%%%%%%%%%%%%%%%%%%%%%%%%%%%%%%%%%%%%%%%%%%%%%%%%%%%%%%%%%%%%%%%%%%%%%%%%%%

We started our analysis from the results of \citet{Xu:2012}, where the dataset, consisting of 
1725 MW and THz frequencies together with 18366 FIR transitions, ranging up to $\varv_t = 2$ 
and $J_{\rm max}$ = 30 for MW/THz and 40 for FIR, was fit using 78 parameters of the RAM 
Hamiltonian, and a weighted standard deviation of 1.071 was achieved. As the first step, 
we have refit this dataset with the RAM36 program \citep{Ilyushin:2010,Ilyushin:2013}. 
Due to the treatment of blends in the RAM36 program, where an intensity-weighted average of 
calculated (but experimentally unresolved) transition frequencies is put in correspondence 
with the measured blended-line frequency, we obtained a slightly better (1.051 here versus 
1.071 in \citet{Xu:2012}) weighted rms deviation for the same data and parameter sets as 
in \citet{Xu:2012}. Thus, in contrast to our initial guesses, the line blending issue 
did not pose a significant problem to the previous fitting attempt \citep{Xu:2012}.

%%%%%%%%%%%%%%%%%%%%%%%%%%%%%%%%%%%%%%%%%%%%%%%%%%%%%%%%%%%%%%%%%%%%%%%%%%%%%%%%%%%%%%%%%
%%%%%%%%%%%%%%%%%%%%%%%%%%%%%%%%%%%%%%%%%%%%%%%%%%%%%%%%%%%%%%%%%%%%%%%%%%%%%%%%%%%%%%%%%

\begin{figure}[t]
 \centering
 \includegraphics[height=6cm]{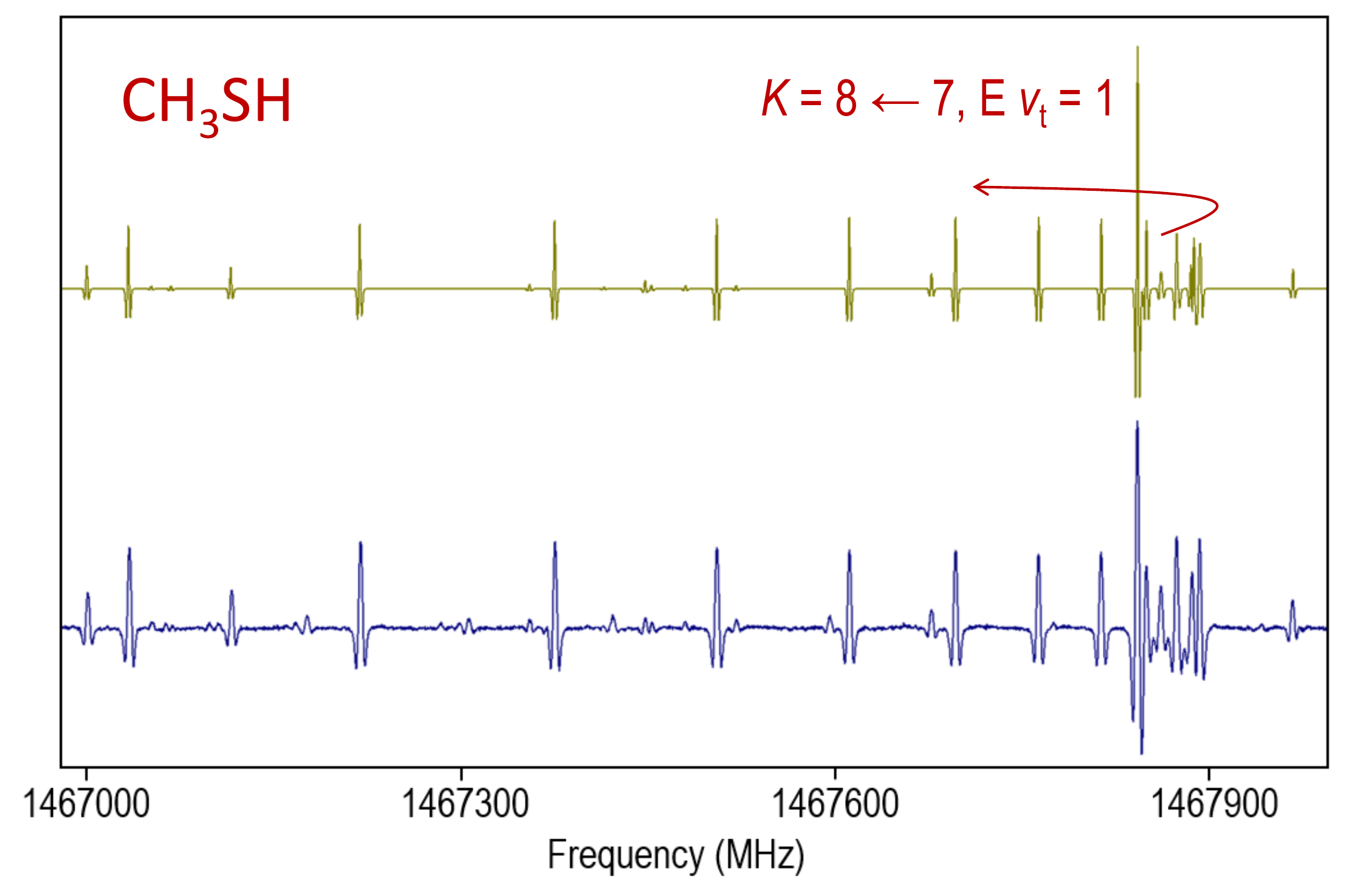}
 \caption{Detail of the CH$_3$SH rotational spectrum showing $b$-type $Q$-branch transitions 
  with $K =8 \leftarrow 7$ for the $\varv_t = 1$ torsional state. The experimental spectrum 
  is shown in the lower trace. A simulation of CH$_3$SH transitions up to $\varv_t = 2$ 
  is shown in the upper trace.}
 \label{fgr:THzSpektrum}
\end{figure}

%%%%%%%%%%%%%%%%%%%%%%%%%%%%%%%%%%%%%%%%%%%%%%%%%%%%%%%%%%%%%%%%%%%%%%%%%%%%%%%%%%%%%%%%%
%%%%%%%%%%%%%%%%%%%%%%%%%%%%%%%%%%%%%%%%%%%%%%%%%%%%%%%%%%%%%%%%%%%%%%%%%%%%%%%%%%%%%%%%%

The new data were assigned starting from the Cologne measurements in the 155$-$510~GHz frequency 
range. The THz records (1.1$-$1.8~THz) were reanalaysed subsequently, based on our new results. 
The Kharkiv measurements in the 49$-$149~GHz range were assigned at the final stage. 
The assignments were done in parallel for all three torsional states under consideration 
since the previous study \citep{Xu:2012} provided rather good starting predictions. 
Whenever it was possible, we have replaced the old measurements (see \citet{Xu:2012} 
and references therein) with the new, more accurate ones. 
In parallel with the assignment process, a search of the optimal set of RAM torsion–rotation parameters 
was fulfilled, in which different parameters up to $n_{\rm op} = 12$ order were tested (the ordering 
scheme of \citet{NAKAGAWA:1987} is assumed). In the process of model refinement, we were able to include 
in the fit the majority of the transitions which were tentatively assigned in \citet{Xu:2012}, 
but not included in the fit due to large residuals between measured and calculated transition frequencies. 
We had to change the assignments only for a very small part of the tentatively assigned lines, mainly 
for some high $J$ transitions which were out of consideration in the previous study \citep{Xu:2012}.

It should be noted that we concentrated mainly on the MW part of the spectrum in our current work, 
keeping in the fit the same set of FIR lines as in \citet{Xu:2012} with the only exception, that part 
of the FIR measurements were replaced with the new, more accurate sub-mm and THz measurements. 
Moreover, we were able to assign the same measurement uncertainty of 0.0002~cm$^{-1}$ for all 
FIR measurements included in the fit. This uncertainty for all of the FIR transitions may even 
seem slightly conservative as we were able to reproduce the FIR data to about 0.00012~cm$^{-1}$ 
on average. In the previous study by \citet{Xu:2012}, uncertainties of 0.00020~cm$^{-1}$ were 
assigned to all infrared transitions from the $\varv_t = 0$ state, and uncertainties of 
0.00035~cm$^{-1}$ were assigned to all infrared transitions from the $\varv_t \geq 1$ states.

The final dataset treated in this work involves 6965 MW and 16345 FIR line frequencies that, 
due to blending, correspond to 27279 transitions with $J_{\rm max} = 61$. Transitions within and 
between $\varv_t = 0$, 1, 2 torsional states are included in the dataset. A fit achieving 
a weighted rms deviation of 0.72 for this dataset with 124 parameters included in the model 
was chosen as our “best fit” for this paper. The 124 molecular parameters obtained from this fit 
are given in the Appendix in Table~\ref{tbl:ParametersTable}. The low order parameters up to 
fourth order may be found in Table~\ref{tbl:ParametersComparison} where they are compared with 
previous results from \citet{Xu:2012}. Despite the large number of parameters, the final fit 
converged perfectly in all three senses: (i) the relative change in the wrms deviation of the fit 
at the last iteration is less than 10$^{-7}$; (ii) the corrections to the parameter values 
generated at the last iteration are less than 10$^{-4}$ of the calculated parameter confidence 
intervals; (iii) the changes generated at the last iteration in the calculated frequencies 
are less than 1~kHz even for the intertorsional FIR transitions. The numbers of the terms 
in the model distributed between the $n_{\rm op} = 2$, 4, 6, 8, 10, 12 orders are 7, 22, 42, 
39, 11, 3 respectively. These values are equal to or less than the total numbers of determinable 
parameters of 7, 22, 50, 95, 161, and 252 for those orders, as calculated from the differences 
between the total number of symmetry-allowed Hamiltonian terms of order $n_{\rm op}$ 
and the number of symmetry-allowed contact transformation terms of order 
$n_{\rm op} - 1$ \citep{NAKAGAWA:1987}.

The quality of the fit chosen as our best fit for this paper can be seen in 
Table~\ref{tbl:statisticInf}. The overall weighted rms deviation is 0.72. 
For the final dataset, the difference between the fits with and without treatment 
of blends was quite significant (weighted rms 0.72 versus 1.08 respectively). 
Part of this difference is due to accidental blending of lines and part is from 
the clustering of A symmetry transitions in the spectrum of methyl mercaptan. 
The fact that all data groups are fit within their experimental uncertainties (see left part
of Table~\ref{tbl:statisticInf} where the data are grouped by measurement uncertainty) seems 
to us completely satisfactory. At the same time, it should be noted that for the MW part 
of the $\varv_t = 2$ torsional state data the fit still gives a weighted rms deviation above 1.0. 
Nevertheless, even for this group of data we have achieved a significant progress in comparison 
with the previous study \citep{Xu:2012} reducing the wrms from 4.369 to 1.3 
(see Table~\ref{tbl:statisticInf}). We also reduced the wrms deviation for the FIR data 
(down to 0.6) which looks more impressive if one takes into account the fact that we merged the 
two 0.00020~cm$^{-1}$ and 0.00035~cm$^{-1}$ uncertainty groups of data into one 0.00020~cm$^{-1}$ 
uncertainty group. We also note that our new measurements provide a higher level 
of accuracy to test the model in comparison with \citet{Xu:2012}. Whereas the most precise group 
of measurements in \citet{Xu:2012} has uncertainties of 0.050~MHz, we have here groups of data 
with 0.010~MHz and 0.020~MHz uncertainties comparable in size, which are fit within 
experimental error. Thus, we can conclude that significant progress in fitting 
the methyl mercaptan spectrum in the lowest three torsional states was achieved.

Figs.~\ref{fgr:mmWaveSpektrum} and \ref{fgr:THzSpektrum} illustrate our current understanding 
of the methyl mercaptan spectrum around 302.7~GHz and 1.467~THz in which observed and predicted spectra 
with our current model are compared. In Fig.~\ref{fgr:mmWaveSpektrum}, a region which is dominated by 
$R$ series of $J = 12 \leftarrow 11$ $a$-type transitions is shown. In Fig.~\ref{fgr:THzSpektrum}, 
a region dominated by $b$-type $Q$-branch transitions with $K = 8 \leftarrow 7$ for the $\varv_t = 1$ 
torsional state is given. A slight inconsistency in intensity between the predicted and the observed spectrum, 
that may be visible between some groups of lines, especially in Fig.~\ref{fgr:mmWaveSpektrum}, is due 
to source power and detector sensitivity variations. It is seen that the majority of strong lines are 
assigned and well predicted by our current model, although a number of unassigned lines, 
presumably belonging to higher excited states or minor isotopic species, are visible 
in the experimental spectrum.

%%%%%%%%%%%%%%%%%%%%%%%%%%%%%%%%%%%%%%%%%%%%%%%%%%%%%%%%%%%%%%%%%%%%%%%%%%%%%%%%%%%%%%%%%
%%%%%%%%%%%%%%%%%%%%%%%%%%%%%%%%%%%%%%%%%%%%%%%%%%%%%%%%%%%%%%%%%%%%%%%%%%%%%%%%%%%%%%%%%
\section{Discussion}
\label{discussion}
%%%%%%%%%%%%%%%%%%%%%%%%%%%%%%%%%%%%%%%%%%%%%%%%%%%%%%%%%%%%%%%%%%%%%%%%%%%%%%%%%%%%%%%%%
%%%%%%%%%%%%%%%%%%%%%%%%%%%%%%%%%%%%%%%%%%%%%%%%%%%%%%%%%%%%%%%%%%%%%%%%%%%%%%%%%%%%%%%%%

We compared our current results with the parameters of the previous study \citep{Xu:2012} 
to have a more detailed picture of how the dataset extension affects the low order parameters in 
the RAM Hamiltonian model of CH$_3$SH. In view of rather large differences in datasets and 
sets of high order torsion-rotational parameters, we limit the comparison of parameters 
up to fourth order only in Table~\ref{tbl:ParametersComparison}. It should be noted that 
some parameter and operator expressions in Table~\ref{tbl:ParametersComparison} 
have changed in comparison with Table~\ref{tbl:ParametersTable}. 
This is caused by the fact that the general Hamiltonian form (1), which is encoded in 
the RAM36 program, does not allow modification of the coefficient in front of the expression. 
Thus all coefficients historically adopted for a number of terms (such as minus sign in front 
of the quartic centrifugal distortion terms) are absorbed in the parameter values. 
For the purpose of the current comparison, we have recalculated all differing parameters 
to the form which conforms with results of \citet{Xu:2012}. 
As we can see from Table~\ref{tbl:ParametersComparison}, there are no big changes either 
in the rotational constants or in the main internal rotational parameters $V_3$, $\rho$, 
and $F$. In addition, many torsion-rotation distortion parameters of the fourth order 
agree well with previous results of \citet{Xu:2012}. The most noticeable differences 
are observed for $\Delta_{JK}$, $F_{bc}$, and $D_{3bc}$, for which the signs changed. 
On one hand, this may be caused by the difference in the fourth order parameter sets. 
Indeed, in the current Hamiltonian model we do not use $D_{abJ}$ and $D_{abK}$, which 
are present in the \citet{Xu:2012} Hamiltonian model. On the other hand, this may be 
a consequence of additional correlation problems in the \citet{Xu:2012} parameter set, 
where the number of fourth order parameters exceeds by 2 the maximum number of 
determinable parameters for this order as predicted by the reduction scheme proposed 
by \citet{NAKAGAWA:1987}. In any case, looking at the rather good agreement between many 
torsion-rotation distortion parameters of fourth order, it seems unlikely that the sign 
changes in $\Delta_{JK}$, $F_{bc}$, and $D_{3bc}$ are caused by the difference in the datasets.

One more issue, which should be discussed in connection with the current parameter set, is 
the expansion of the torsional potential function. It is seen from the comparison of the 
$V_3$, $V_6$, $V_9$, $V_{12}$ values that our expansion of the potential function is 
far from smooth convergence. In the previous study \citep{Xu:2012}, the convergence 
of the potential function expansion raised only minor suspicion since \textit{$V_6$} 
and $V_9$ were of the same order ($-$0.572786(15)~cm$^{-1}$ and 0.205603(31)~cm$^{-1}$, 
respectively), whereas problems in the expansion convergence are more obvious in 
the present study because $V_9$ is larger than $V_6$, and $V_{12}$ is larger than $V_9$. 
It is known that the expansion coefficients of the torsional potential function 
may be highly correlated with Fermi-type couplings with small amplitude vibrations 
\citep{Moazzen-Ahmadi:2002, Gascooke:2015}. Thus, the current potential function 
behavior may be explained by the intervibrational interactions with the low lying 
small amplitude vibrational modes in methyl mercaptan. Indeed, the study of \citet{Lees:2016} 
revealed strong torsion–vibrational resonant coupling between the $\varv_t = 4$ torsional 
state and the CS stretching vibrational state of methyl mercaptan. 
The higher values of $J$ and $K$ accessed in the present study may give rise to a 
larger amount of perturbations between small amplitude vibrations and higher excited 
torsional states, and these perturbations can be transfered down to lower excited 
torsional states through intertorsional interactions. 
Our first attempts to include in the Hamiltonian model explicit interactions with 
low lying vibrational states lead to significant reductions in the $V_9$ and $V_{12}$ 
values, thus supporting the explanation above. This analysis of MW, FIR, and mid-IR 
spectra of the CS stretch state of methyl mercaptan are ongoing, and results will be 
presented elsewhere in due course.

%%%%%%%%%%%%%%%%%%%%%%%%%%%%%%%%%%%%%%%%%%%%%%%%%%%%%%%%%%%%%%%%%%%%%%%%%%%%%%%%%%%%%%%%%
%%%%%%%%%%%%%%%%%%%%%%%%%%%%%%%%%%%%%%%%%%%%%%%%%%%%%%%%%%%%%%%%%%%%%%%%%%%%%%%%%%%%%%%%%
\section{Spectroscopic database}
\label{database}
%%%%%%%%%%%%%%%%%%%%%%%%%%%%%%%%%%%%%%%%%%%%%%%%%%%%%%%%%%%%%%%%%%%%%%%%%%%%%%%%%%%%%%%%%
%%%%%%%%%%%%%%%%%%%%%%%%%%%%%%%%%%%%%%%%%%%%%%%%%%%%%%%%%%%%%%%%%%%%%%%%%%%%%%%%%%%%%%%%%

One outcome of the present work is a list of transitions calculated from the parameters 
of our final fit. This list includes information on transition quantum numbers, transition 
frequencies, calculated uncertainties, lower state energies, and transition strengths. 
Since extrapolation beyond the quantum number coverage of any given measured dataset 
rapidly becomes unreliable, especially in the case of molecules with large amplitude motions, 
we have chosen a torsional state limit of $\varv_t \leq 2$ and rotational limits of 
$J \leq 70$ and $|K_{a}| \leq 20$. As it was already mentioned, we label 
torsion-rotation levels by the free rotor quantum number $m$, the overall rotational 
angular momentum quantum number $J$, and a signed value of $K_a$. For convenience, 
a $K_c$ value is also given but it is simply recalculated from $J$ and $K_{a}$ values 
($K_{c} = J - |K_{a}|$ for $K_{a} \geq 0$ and $K_{c} = J - |K_{a}|$ + 1 for $K_{a}$ < 0). 
The $m = 0$, $-$3, 3 / 1, $-$2, 4 values correspond to A/E transitions of the $\varv_t = 0$, 
1, 2 torsional states respectively. The predictions range up to 2~THz, and we limit 
our predictions to transitions with uncertainties less than 0.1~MHz. Lower state energies 
are given, referenced to the $J = 0$ A-type $\varv_t = 0$ level. 
This level was calculated to be 107.49563~cm$^{-1}$ above the bottom 
of the torsional potential well. The line strengths in the present line list were calculated 
using the values $\mu_a=1.312$~D and $\mu_b=-0.758$~D \citep{TSUNEKAWA:1989}, which were 
recalculated to the RAM axis system of the current study. In addition, we provide 
the rotation-torsion part of the partition function $Q_{rt}(T)$ of methyl mercaptan 
calculated from first principles (Table \ref{tbl:TabQ}), that is, via direct summation 
over the rotational-torsion states. The maximum value of the $J$ quantum number for 
the energy levels taken for calculating the partition function is 90 and $n_{vt}=11$ 
torsional states were taken into account. The predictions, as well as line list 
of the data set treated in the present work, may be found in the online 
supplementary material with this article.

%%%%%%%%%%%%%%%%%%%%%%%%%%%%%%%%%%%%%%%%%%%%%%%%%%%%%%%%%%%%%%%%%%%%%%%%%%%%%%%%%%%%%%%%%
%%%%%%%%%%%%%%%%%%%%%%%%%%%%%%%%%%%%%%%%%%%%%%%%%%%%%%%%%%%%%%%%%%%%%%%%%%%%%%%%%%%%%%%%%
\section{Conclusions}
\label{conclusion}
%%%%%%%%%%%%%%%%%%%%%%%%%%%%%%%%%%%%%%%%%%%%%%%%%%%%%%%%%%%%%%%%%%%%%%%%%%%%%%%%%%%%%%%%%
%%%%%%%%%%%%%%%%%%%%%%%%%%%%%%%%%%%%%%%%%%%%%%%%%%%%%%%%%%%%%%%%%%%%%%%%%%%%%%%%%%%%%%%%%

We have presented a new study of rotational spectra in the lowest three torsional states of 
methyl mercaptan main isotoplog CH$_3$$^{32}$SH using a rotation–torsion RAM Hamiltonian. 
In the current study, the dataset available in the literature was augmented by new measurements 
in the 49$-$510~GHz range as well as new assignments in the 1.1$-$1.8~THz range. 
The set of 124 RAM Hamiltonian parameters fit with a weighted rms deviation of 0.72  
the data set of 6965 MW and 16345 FIR line frequencies, which sample 
both A and E species of the $\varv_t = 0$, 1, 2 torsional states with $J \leq 61$ and 
$K_{a} \leq 18$ and which cover the frequency range from 7~GHz to 1.8~THz for MW 
lines and up to 482~cm$^{-1}$ for FIR lines. Based on these results, reliable frequency 
predictions were produced for astrophysical use up to 2~THz. 
These are available as supplementary material to this article. The predictions, 
as well as other supplementary files, will also be available in the Cologne Database 
for Molecular Spectroscopy\footnote{https://cdms.astro.uni-koeln.de/classic/entries/; 
https://cdms.astro.uni-koeln.de/classic/predictions/daten/Methanethiol/}, CDMS \citep{CDMS_3}.

\begin{acknowledgements}  
The present study was supported by the Deutsche Forschungsgemeinschaft (DFG) in the 
framework of the collaborative research grant SFB 956 (project ID 184018867), 
sub-project B3. O.Z. is funded by the DFG via the Ger{\"a}tezentrum 
"Cologne Center for Terahertz Spectroscopy" (project ID SCHL 341/15-1). 
The research in Kharkiv was carried out under support of the Volkswagen foundation. 
The assistance of the Science and Technology Center in the Ukraine is acknowledged 
(STCU partner project P686). L.H.X. and R.M.L. received support from the Natural 
Sciences and Engineering Research Council of Canada.
\end{acknowledgements}

%%%%%%%%%%%%%%%%%%%%%%%%%%%%%%%%%%%%%%%%%%%%%%%%%%%%%%%%%%%%%%%%%%%%%%%%%%%%%%%%%%%%%%%%%
%%%%%%%%%%%%%%%%%%%%%%%%%%%%%%%%%%%%%%%%%%%%%%%%%%%%%%%%%%%%%%%%%%%%%%%%%%%%%%%%%%%%%%%%%
 
\bibliographystyle{aa} % style aa.bst
\bibliography{bibliography} % your references Yourfile.bib

%%%%%%%%%%%%%%%%%%%%%%%%%%%%%%%%%%%%%%%%%%%%%%%%%%%%%%%%%%%%%%%%%%%%%%%%%%%%%%%%%%%%%%%%%%
%%%%%%%%%%%%%%%%%%%%%%%%%%%%%%%%%%%%%%%%%%%%%%%%%%%%%%%%%%%%%%%%%%%%%%%%%%%%%%%%%%%%%%%%%%

%%%%%%%%%%%%%%%%%%%%%%%%%%%%%%%%%%%%%%%%%%%%%%%%%%%%%%%%%%%%%%%%%%%%%
%%%%%%%%%%%%%%%%%%%%%%%%%%%%%%%%%%%%%%%%%%%%%%%%%%%%%%%%%%%%%%%%%%%%%
%%%%%%    Table 1    %%%%%%%%%%%%%%%%%%%%%%%%%%%%%%%%%%%%%%%%%%%%%%%%
%%%%%%%%%%%%%%%%%%%%%%%%%%%%%%%%%%%%%%%%%%%%%%%%%%%%%%%%%%%%%%%%%%%%%
%%%%%%%%%%%%%%%%%%%%%%%%%%%%%%%%%%%%%%%%%%%%%%%%%%%%%%%%%%%%%%%%%%%%%

\longtab{
\begin{longtable}{llllll}
\caption{\label{tbl:ParametersComparison} Comparison of selected fit CH$_3$$^{32}$SH parameters with previous results}\\
\hline\hline
$n_{tr}$$^a$ & Operator$^b$ & Par.$^{c}$ & Current work$^{d,e}$ & \citet{Xu:2012} $^{d,e}$ \\
\hline
\endfirsthead
\caption{continued.}\\
\hline\hline
$n_{tr}$$^a$ & Operator$^b$ & Par.$^{c}$ & Current work$^{d,e}$ & \citet{Xu:2012} $^{d,e}$  \\
\hline
\endhead
\hline
    $2_{2,0}$ & $p_\alpha^2$               & $F$ & 15.04062399(54) & 15.04020465(66) \\
    $2_{2,0}$ & $(1/2)(1-\cos 3\alpha)$    & $V_3$ & 441.69136(24) & 441.442236(10) \\
    $2_{1,1}$ & $p_\alpha P_a$             & $\rho$ & 0.6518557764(11) & 0.651856026(13) \\
    $2_{0,2}$ & $P_a^2$                    & $A$ & 3.4279249(17) & 3.42808445(84) \\
    $2_{0,2}$ & $P_b^2$                    & $B$ & 0.4320294(32) & 0.43201954(87) \\
    $2_{0,2}$ & $P_c^2$                    & $C$ & 0.4132203(21) & 0.41325076(83) \\
    $2_{0,2}$ & $\{P_a{,}P_b\}$            & $D_{ab}$ & $-0.00737202(42)$ & $-0.0073126(59)$ \\
    $4_{4,0}$ & $(1/2)(1-\cos 6\alpha)$    & $V_6$ & $-1.9212(12)$ & $-0.572786(15)$ \\ 
    $4_{4,0}$ & $p_\alpha ^4$              & $F_m$ & $-0.1121789(22)\times 10^{-2}$ & $-0.114016(10)\times 10^{-2}$ \\
    $4_{3,1}$ & $p_\alpha ^3 P_a$          & $\rho_m$ & $-0.3554280(60)\times 10^{-2}$ & $-0.360009(28)\times 10^{-2}$ \\
    $4_{2,2}$ & $P^2(1-\cos 3\alpha)$      & $V_{3J}$ & $-0.2062768(53)\times 10^{-2}$ & $-0.217540(84)\times 10^{-2}$ \\
    $4_{2,2}$ & $P_a^2 (1-\cos 3\alpha)$   & $V_{3K}$ & $0.7277626(39)\times 10^{-2}$ & $0.724978(19)\times 10^{-2}$ \\ 
    $4_{2,2}$ & $(P_b^2-P_c^2)(1-\cos 3\alpha)$ & $V_{3bc}$ & $-0.81043(38)\times 10^{-4}$ & $-0.92104(47)\times 10^{-4}$ \\
    $4_{2,2}$ & $\{P_a{,}P_b\}(1-\cos 3\alpha)$ & $V_{3ab}$ & $0.614197(19)\times 10^{-2}$ & $0.61562(30)\times 10^{-2}$ \\  
    $4_{2,2}$ & $p^2_{\alpha} P^2$         & $F_J$ & $-0.3094800(28)\times 10^{-4}$ & $-0.8106(38)\times 10^{-4}$ \\
    $4_{2,2}$ & $p^2_{\alpha} P_a^2$       & $F_K$ & $-0.4789751(62)\times 10^{-2}$ & $-0.483287(30)\times 10^{-2}$ \\  
    $4_{2,2}$ & $p_\alpha ^2\{P_a{,}P_b\}$ & $F_{ab}$ & $0.10745(45)\times 10^{-4}$ & $0.843(45)\times 10^{-4}$ \\
    $4_{2,2}$ & $2p^2_{\alpha}(P_b^2-P_c^2)$ & $F_{bc}$ & $-0.32942(41)\times 10^{-4}$ & $0.0536(41)\times 10^{-4}$ \\
    $4_{2,2}$ & $\{P_a{,}P_c\}\sin 3\alpha$ & $D_{3ac}$ & $0.077299(15)\times 10^{-1}$ & $0.1036(15)\times 10^{-1}$ \\
    $4_{2,2}$ & $\{P_b{,}P_c\}\sin 3\alpha$ & $D_{3bc}$ & $-0.6418(14)\times 10^{-3}$ & $0.665(14)\times 10^{-3}$ \\  
    $4_{1,3}$ & $p_\alpha P_a P^2$         & $\rho_J$ & $-0.4255507(38)\times 10^{-4}$ & $-0.4726(54)\times 10^{-4}$ \\
    $4_{1,3}$ & $p_\alpha P_a^3$           & $\rho_K$ & $-0.2958211(29)\times 10^{-2}$ & $-0.30381(74)\times 10^{-2}$ \\
    $4_{1,3}$ & $p_\alpha \{P_a^2{,}P_b\}$ & $\rho_{ab}$ & $0.10025(43)\times 10^{-4}$ & $0.999(67)\times 10^{-4}$ \\  
    $4_{1,3}$ & $p_\alpha \{P_a{,}(P_b^2-P_c^2)\}$ & $\rho_{bc}$ & $-0.42674(40)\times 10^{-4}$ & $-0.0462(39)\times 10^{-4}$ \\
    $4_{0,4}$ & $-P^4$                     & $\Delta_J$ & $0.5393457(89)\times 10^{-6}$ & $0.538140(23)\times 10^{-6}$ \\
    $4_{0,4}$ & $-P^2 P_a^2$               & $\Delta_{JK}$ & $0.1784933(23)\times 10^{-4}$ & $-0.066(26)\times 10^{-5}$ \\
    $4_{0,4}$ & $-P_a^4$                   & $\Delta_K$ & $0.6990318(56)\times 10^{-3}$ & $0.7425(48)\times 10^{-3}$ \\
    $4_{0,4}$ & $-2P^2(P_b^2-P_c^2)$       & $\delta_J$ & $0.2281975(75)\times 10^{-7}$ & $0.224788(88)\times 10^{-7}$ \\
    $4_{0,4}$ & $-\{P_a^2{,}(P_b^2-P_c^2)\}$ & $\delta_K$ & $0.109149(20)\times 10^{-4}$ & $0.10483(32)\times 10^{-4}$ \\
    $4_{0,4}$ & $P^2\{P_a{,}P_b\}$         & $D_{abJ}$ & $ - $  & $-0.956(60)\times 10^{-7}$ \\    
    $4_{0,4}$ & $\{P_a^3{,}P_b\}$          & $D_{abK}$ & $ - $ & $0.202(23)\times 10^{-4}$ \\    
     &                                     & $\theta_{\rm RAM}$ & $-$0.14$^\circ$ & $-$0.14$^\circ$ \\      
\hline
\end{longtable}
\tablefoot{$^{a}$ $n=t+r$, where $n$ is the total order of the operator, $t$ is the order of the torsional part 
and $r$ is the order of the rotational part, respectively. The ordering scheme of \citet{NAKAGAWA:1987} is used. 
$^{b}$ \{A,B\} = AB + BA. The product of the operator in the second column of a given row and the parameter 
in the third column of that row gives the term actually used in the torsion-rotation Hamiltonian of the program, 
except for $F$, $\rho$ and $A_{\rm RAM}$, which occur in the Hamiltonian in the form 
$F(p_a + \rho P_a)^2 + A_{\rm RAM}P_a^2$. $^{c}$ The parameter nomenclature is based on the subscript 
procedure of \citet{XU:2008305}.  $^{d}$ Values of the parameters in cm$^{-1}$, except for $\rho$, which is unitless, 
and for $\theta_{\rm RAM}$, which is in degrees. $^{e}$ Statistical uncertainties are given in parentheses as one standard 
uncertainty in units of the last digits.}
}

%%%%%%%%%%%%%%%%%%%%%%%%%%%%%%%%%%%%%%%%%%%%%%%%%%%%%%%%%%%%%%%%%%%%%%%%%%%%%%%%%%%%%%%%%%
%%%%%%%%%%%%%%%%%%%%%%%%%%%%%%%%%%%%%%%%%%%%%%%%%%%%%%%%%%%%%%%%%%%%%%%%%%%%%%%%%%%%%%%%%%

%%%%%%%%%%%%%%%%%%%%%%%%%%%%%%%%%%%%%%%%%%%%%%%%%%%%%%%%%%%%%%%%%%%%%
%%%%%%%%%%%%%%%%%%%%%%%%%%%%%%%%%%%%%%%%%%%%%%%%%%%%%%%%%%%%%%%%%%%%%
%%%%%%    Table 2    %%%%%%%%%%%%%%%%%%%%%%%%%%%%%%%%%%%%%%%%%%%%%%%%
%%%%%%%%%%%%%%%%%%%%%%%%%%%%%%%%%%%%%%%%%%%%%%%%%%%%%%%%%%%%%%%%%%%%%
%%%%%%%%%%%%%%%%%%%%%%%%%%%%%%%%%%%%%%%%%%%%%%%%%%%%%%%%%%%%%%%%%%%%%

\begin{table*}[]
\caption{\label{tbl:statisticInf} Overview of the data set and fit quality }
\begin{tabular}{lll|lll|lll}
\hline
\multicolumn{3}{c|}{By measurement uncertainty}                                           & \multicolumn{6}{c}{By torsional state}                                                                                                                                                      \\ \cline{4-9} 
\multicolumn{3}{l|}{}                                                                     & \multicolumn{3}{c|}{MW data}                                                           & \multicolumn{3}{c}{FIR data}                                                              \\ \hline
\multicolumn{1}{c}{Unc.$^a$} & \multicolumn{1}{c}{$\#^b$} & \multicolumn{1}{c|}{rms$^c$} & \multicolumn{1}{c}{$\varv_t^d$} & \multicolumn{1}{c}{$\#^b$} & \multicolumn{1}{c|}{wrms$^e$} & \multicolumn{1}{c}{$\varv_t^d$} & \multicolumn{1}{c}{$\#^b$} & \multicolumn{1}{c}{wrms$^e$} \\ \hline
0.010~MHz                    & 703                        & 0.0087~MHz                    & $\varv_t=0 \leftarrow 0$       & 3551                       & 0.68(2.075)                    & $\varv_t=0 \leftarrow 0$       & 1436                       & 0.54(0.640)                   \\
0.020~MHz                    & 3501                       & 0.0187~MHz                    & $\varv_t=1 \leftarrow 1$       & 2618                       & 0.93(2.102)                    & $\varv_t=1 \leftarrow 0$       & 8537                       & 0.44(0.785)                   \\
0.030~MHz                    & 162                        & 0.0240~MHz                    & $\varv_t=2 \leftarrow 2$       & 1617                       & 1.35(4.369)                    & $\varv_t=1 \leftarrow 1$       & 896                        & 0.75(1.139)*                  \\
0.050~MHz                    & 1307                       & 0.0495~MHz                    & $\varv_t=2 \leftarrow 1$       & 27                         & 1.87                           & $\varv_t=2 \leftarrow 0$       & 1129                       & 0.72(1.195)                   \\
0.100~MHz                    & 613                        & 0.0923~MHz                    &                                &                            &                                & $\varv_t=2 \leftarrow 1$       & 6573                       & 0.67(1.283)*                  \\
0.200~MHz                    & 679                        & 0.1892~MHz                    &                                &                            &                                & $\varv_t=2 \leftarrow 2$       & 895                        & 0.79(1.825)*                  \\
$2\times 10^{-4}$~cm$^{-1}$  & 16345                      & $1.2\times 10^{-4}$~cm$^{-1}$ &                                &                            &                                &                                &                            &                               \\ \hline
\end{tabular}
\tablefoot{$^{a}$ Estimated measurement uncertainties for each data group. $^{b}$ Number of lines (left part) or transitions (middle and right parts) 
of each category in the least-squares fit. Due to blending, 27279 transitions correspond to 23310 measured line frequencies in the fit. 
$^{c}$ Root-mean-square (rms) deviation of corresponding data group. $^{d}$ Upper and lower state torsional quantum number $\varv_t$. 
$^{e}$ Weighted root-mean-square (wrms) deviation of corresponding data group. The corresponding value from the previous work of \citet{XU:2008305} 
is given in parentheses. For categories marked with an asterisk, the wrms deviation was recalculated taking into account that in the current work 
for these categories we changed the uncertainty from $3.5\times 10^{-4}$~cm$^{-1}$ to $2.0\times 10^{-4}$~cm$^{-1}$.}
\end{table*}

%%%%%%%%%%%%%%%%%%%%%%%%%%%%%%%%%%%%%%%%%%%%%%%%%%%%%%%%%%%%%%%%%%%%%%%%%%%%%%%%%%%%%%%%%
%%%%%%%%%%%%%%%%%%%%%%%%%%%%%%%%%%%%%%%%%%%%%%%%%%%%%%%%%%%%%%%%%%%%%%%%%%%%%%%%%%%%%%%%%

%%%%%%%%%%%%%%%%%%%%%%%%%%%%%%%%%%%%%%%%%%%%%%%%%%%%%%%%%%%%%%%%%%%%%
%%%%%%%%%%%%%%%%%%%%%%%%%%%%%%%%%%%%%%%%%%%%%%%%%%%%%%%%%%%%%%%%%%%%%
%%%%%%    Table 3    %%%%%%%%%%%%%%%%%%%%%%%%%%%%%%%%%%%%%%%%%%%%%%%%
%%%%%%%%%%%%%%%%%%%%%%%%%%%%%%%%%%%%%%%%%%%%%%%%%%%%%%%%%%%%%%%%%%%%%
%%%%%%%%%%%%%%%%%%%%%%%%%%%%%%%%%%%%%%%%%%%%%%%%%%%%%%%%%%%%%%%%%%%%%

\begin{table}
\centering
\caption{\label{tbl:TabQ} Rotation-torsion part $Q_{rt}(T)$ of the total partition function calculated 
 from first principles using the parameter set of Table \ref{tbl:ParametersTable}.}
\begin{tabular}{cc}
\hline\hline
Temperature(K) & $Q_{rt}(T)$ \\
\hline
300   & 22795.4 \\
225   & 12582.3 \\
200   & 9933.7  \\
150   & 5703.62 \\
100   & 2774.0  \\
75    & 1729.89 \\
37.5  & 592.634 \\
18.57 & 205.742 \\
9.375 & 70.5870 \\
5.000 & 26.2970 \\
2.725 & 10.0228 \\
\hline
\end{tabular}
\end{table}

\begin{appendix}

\section{Complementary table}
\label{appendix_table}

Table~\ref{tbl:ParametersTable} summarizes the full set of 124 
spectroscopic parameters determined in the present study.

%%%%%%%%%%%%%%%%%%%%%%%%%%%%%%%%%%%%%%%%%%%%%%%%%%%%%%%%%%%%%%%%%%%%%
%%%%%%%%%%%%%%%%%%%%%%%%%%%%%%%%%%%%%%%%%%%%%%%%%%%%%%%%%%%%%%%%%%%%%
%%%%%%    Table A1   %%%%%%%%%%%%%%%%%%%%%%%%%%%%%%%%%%%%%%%%%%%%%%%%
%%%%%%%%%%%%%%%%%%%%%%%%%%%%%%%%%%%%%%%%%%%%%%%%%%%%%%%%%%%%%%%%%%%%%
%%%%%%%%%%%%%%%%%%%%%%%%%%%%%%%%%%%%%%%%%%%%%%%%%%%%%%%%%%%%%%%%%%%%%

\longtab[1]{
\begin{longtable}{llll}
\caption{\label{tbl:ParametersTable} Fit parameters of the RAM Hamiltonian for CH$_3$$^{32}$SH molecule}\\
\hline\hline
$n_{tr}$$^a$ & Operator$^b$ & Par.$^{c}$ & Value$^{d,e}$ \\
\hline
\endfirsthead
\caption{continued.}\\
\hline\hline
$n_{tr}$$^a$ & Operator$^b$ & Par.$^{c}$ & Value$^{d,e}$ \\
\hline
\endhead
\hline
    $2_{2,0}$ & $p_\alpha^2$                & $F$         & 15.04062399(54)  \\
    $2_{2,0}$ & $(1-\cos 3\alpha)$          & $(1/2)V_3$  & 220.84568(12)  \\
    $2_{1,1}$ & $p_\alpha P_a$              & $\rho$      & 0.6518557764(11)  \\
    $2_{0,2}$ & $P_a^2$                     & $A$  & 3.4279249(17)  \\
    $2_{0,2}$ & $P_b^2$                     & $B$  & 0.4320294(32)  \\
    $2_{0,2}$ & $P_c^2$                     & $C$  & 0.4132203(21)  \\
    $2_{0,2}$ & $(1/2)\{P_a{,}P_b\}$        & $2D_{ab}$   & $-0.01474403(84)$  \\
    $4_{4,0}$ & $p_\alpha ^4$               & $F_m$      & $-0.1121789(22)\times 10^{-2}$  \\
    $4_{4,0}$ & $(1-\cos 6\alpha)$          & $(1/2)V_6$ & $-0.96060(59)$  \\ 
    $4_{3,1}$ & $p_\alpha ^3 P_a$           & $\rho_m$   & $-0.3554280(60)\times 10^{-2}$  \\
    $4_{2,2}$ & $p^2_{\alpha} P^2$          & $F_J$ & $-0.3094800(28)\times 10^{-4}$  \\
    $4_{2,2}$ & $p^2_{\alpha} P_a^2$        & $F_K$ & $-0.4789751(62)\times 10^{-2}$ \\  
    $4_{2,2}$ & $(1/2)p_\alpha ^2\{P_a{,}P_b\}$ & $F_{ab}$ & $0.21490(90)\times 10^{-4}$  \\
    $4_{2,2}$ & $p^2_{\alpha}(P_b^2-P_c^2)$ & $F_{bc}$ & $-0.65884(82)\times 10^{-4}$  \\
    $4_{2,2}$ & $P^2(1-\cos 3\alpha)$       & $V_{3J}$ & $-0.2062768(53)\times 10^{-2}$ \\
    $4_{2,2}$ & $P_a^2(1-\cos 3\alpha)$     & $V_{3K}$ & $0.7277626(39)\times 10^{-2}$  \\ 
    $4_{2,2}$ & $(P_b^2-P_c^2)(1-\cos 3\alpha)$ & $V_{3bc}$ & $-0.81043(38)\times 10^{-4}$  \\
    $4_{2,2}$ & $(1/2)\{P_a{,}P_b\}(1-\cos 3\alpha)$ & $V_{3ab}$ & $0.1228394(37)\times 10^{-1}$  \\  
    $4_{2,2}$ & $(1/2)\{P_a{,}P_c\}\sin 3\alpha$ & $D_{3ac}$ & $0.154598(30)\times 10^{-1}$ \\
    $4_{2,2}$ & $(1/2)\{P_b{,}P_c\}\sin 3\alpha$ & $D_{3bc}$ & $-0.12835(27)\times 10^{-2}$ \\  
    $4_{1,3}$ & $p_\alpha P_aP^2$           & $\rho_J$ & $-0.4255507(38)\times 10^{-4}$ \\
    $4_{1,3}$ & $p_\alpha P_a^3$            & $\rho_K$ & $-0.2958211(29)\times 10^{-2}$ \\
    $4_{1,3}$ & $(1/2)\{P_a{,}(P_b^2-P_c^2)\}p_\alpha$ & $\rho_{bc}$ & $-0.85348(80)\times 10^{-4}$ \\
    $4_{1,3}$ & $(1/2)p_\alpha \{P_a^2{,}P_b\}$  & $\rho_{ab}$ & $0.20049(86)\times 10^{-4}$ \\
    $4_{0,4}$ & $P^4$                       & $-\Delta_J$ & $-0.5393457(89)\times 10^{-6}$ \\
    $4_{0,4}$ & $P^2 P_a^2$                 & $-\Delta_{JK}$ & $-0.1784933(23)\times 10^{-4}$  \\
    $4_{0,4}$ & $P_a^4$                     & $-\Delta_K$ & $-0.6990318(56)\times 10^{-3}$ \\
    $4_{0,4}$ & $P^2(P_b^2-P_c^2)$          & $-2\delta_J$ & $-0.456395(15)\times 10^{-7}$  \\
    $4_{0,4}$ & $(1/2)\{P_a^2{,}(P_b^2-P_c^2)\}$ &  $-2\delta_K$ & $-0.218299(40)\times 10^{-4}$ \\
    $6_{6,0}$ & $p_\alpha ^6$               & $F_{mm}$ & $-0.20877(12)\times 10^{-5}$ \\
    $6_{6,0}$ & $(1-\cos 9\alpha)$          & $(1/2)V_9$ & 2.9920(25) \\   
    $6_{5,1}$ & $p_\alpha ^5 P_a$           & $\rho_{mm}$ & $-0.76728(49)\times 10^{-5}$ \\
    $6_{4,2}$ & $p_\alpha ^4P^2$            & $F_{mJ}$ & $0.3736(27)\times 10^{-8}$ \\    
    $6_{4,2}$ & $p_\alpha ^4P_a^2$          & $F_{mK}$ & $-0.113906(83)\times 10^{-4}$ \\ 
    $6_{4,2}$ & $p^4_{\alpha}(P_b^2-P_c^2)$ & $F_{mbc}$ & $0.3796(76)\times 10^{-7}$  \\
    $6_{4,2}$ & $P^2(1-\cos 6\alpha)$       & $V_{6J}$ & $-0.1915(26)\times 10^{-4}$ \\         
    $6_{4,2}$ & $P_a^2(1-\cos 6\alpha)$     & $V_{6K}$ & $-0.22093(13)\times 10^{-3}$ \\ 
    $6_{4,2}$ & $(P_b^2-P_c^2)(1-\cos 6\alpha)$ & $V_{6bc}$ & $-0.6466(26)\times 10^{-4}$ \\
    $6_{4,2}$ & $(1/2)\{P_a{,}P_c\}\sin 6\alpha$  & $D_{6ac}$ & $0.31308(91)\times 10^{-3}$ \\     
    $6_{3,3}$ & $p_\alpha ^3P_aP^2$         & $\rho_{mJ}$ & $0.8891(70)\times 10^{-8}$ \\   
    $6_{3,3}$ & $p_\alpha ^3P_a^3$          & $\rho_{mK}$ & $-0.85938(76)\times 10^{-5}$ \\               
    $6_{3,3}$ & $(1/2)\{P_a{,}(P_b^2-P_c^2)\}p^3_\alpha$ & $\rho_{mbc}$ & $ 0.1205(21)\times 10^{-6}$ \\
    $6_{2,4}$ & $p^2_{\alpha} P^4$          & $F_{JJ}$ & $0.20615(36)\times 10^{-9}$ \\       
    $6_{2,4}$ & $p^2_{\alpha} P_a^2 P^2$    & $F_{JK}$ & $0.9542(67)\times 10^{-8}$ \\
    $6_{2,4}$ & $p^2_{\alpha} P_a^4$        & $F_{KK}$ & $-0.33234(41)\times 10^{-5}$ \\ 
    $6_{2,4}$ & $p^2_\alpha P^2(P_b^2-P_c^2)$ &  $F_{bcJ}$ & $-0.8182(45)\times 10^{-10}$ \\
    $6_{2,4}$ & $(1/2)p^2_\alpha \{P_a^2{,}(P_b^2-P_c^2)\}$ & $F_{bcK}$ & $0.1480(20)\times 10^{-6}$ \\
    $6_{2,4}$ & $(1/2)\{P_b^2{,}P_c^2\}p^2_{\alpha}$ & $F_{b2c2}$ & $-0.1738(14)\times 10^{-9}$ \\
    $6_{2,4}$ & $P^4(1-\cos 3\alpha)$       & $V_{3JJ}$ & $0.51439(29)\times 10^{-8}$ \\
    $6_{2,4}$ & $P^2P_a^2(1-\cos 3\alpha)$   & $V_{3JK}$ & $-0.26484(14)\times 10^{-6}$ \\
    $6_{2,4}$ & $P_a^4(1-\cos 3\alpha)$     & $V_{3KK}$ & $0.46471(31)\times 10^{-6}$ \\
    $6_{2,4}$ & $(1/2)P^2\{P_a{,}P_b\}(1-\cos 3\alpha)$ & $V_{3abJ}$ & $0.11219(57)\times 10^{-7}$ \\  
    $6_{2,4}$ & $(1/2)\{P_a^3{,}P_b\}(1 - \cos 3\alpha)$ & $V_{3abK}$ & $-0.4746(37)\times 10^{-6}$ \\
    $6_{2,4}$ & $P^2(P_b^2-P_c^2)(1-\cos 3\alpha)$ & $V_{3bcJ}$ & $0.7672(16)\times 10^{-9}$ \\       
    $6_{2,4}$ & $(1/2)\{P_a{,}P_b^3\}\cos 3\alpha$ & $V_{3ab3}$ & $0.15041(20)\times 10^{-6}$  \\  
    $6_{2,4}$ & $(1/2)\{P_b^2{,}P_c^2\}\cos 3\alpha$ & $V_{3b2c2}$ & $0.5263(30)\times 10^{-8}$  \\  
    $6_{2,4}$ & $(1/2)P^2\{P_a{,}P_c\}\sin 3\alpha$ & $D_{3acJ}$ & $-0.28198(38)\times 10^{-6}$  \\  
    $6_{2,4}$ & $(1/2)\{P_a^3{,}P_c\}\sin 3\alpha$ & $D_{3acK}$ & $-0.3880(64)\times 10^{-6}$  \\  
    $6_{2,4}$ & $(1/2)\{P_a{,}P_c^3\}sin 3\alpha$ & $D_{3ac3}$ & $0.16507(25)\times 10^{-6}$ \\
    $6_{2,4}$ & $(1/2)(\{P_b^3{,}P_c\}-\{P_b{,}P_c^3\})\sin 3\alpha$ & $D_{3bcbc}$ & $-0.45717(51)\times 10^{-8}$ \\  
    $6_{1,5}$ & $p_\alpha P_aP^4$           & $\rho_{JJ}$ & $0.22733(31)\times 10^{-9}$ \\
    $6_{1,5}$ & $p_\alpha P_a^3 P^2$        & $\rho_{JK}$ & $0.5427(29)\times 10^{-8}$ \\
    $6_{1,5}$ & $p_\alpha P_a^5$            & $\rho_{KK}$ & $-0.5450(13)\times 10^{-6}$ \\
    $6_{1,5}$ & $(1/2)\{P_a^3{,}(P_b^2-P_c^2)\}p_\alpha$ & $\rho_{bcK}$ & $0.8249(85)\times 10^{-7}$ \\
    $6_{0,6}$ & $P^6$                       & $\Phi_J$ & $-0.25598(61)\times 10^{-12}$ \\
    $6_{0,6}$ & $P^4 P_a^2$                 & $\Phi_{JK}$ & $0.7921(13)\times 10^{-10}$ \\
    $6_{0,6}$ & $P^2P_a^4$                  & $\Phi_{KJ}$ & $0.13245(45)\times 10^{-8}$  \\
    $6_{0,6}$ & $P_a^6$                     & $\Phi_K$ & $-0.916(18)\times 10^{-8}$ \\
    $6_{0,6}$ & $P^4(P_b^2-P_c^2)$         & $2\phi_J$ & $-0.1142(10)\times 10^{-12}$ \\ 
    $6_{0,6}$ & $(1/2)P^2\{P_a^2{,}(P_b^2-P_c^2)\}$ & $2\phi_{JK}$ & $0.9298(40)\times 10^{-10}$ \\
    $6_{0,6}$ & $(1/2)\{P_a^4{,}(P_b^2-P_c^2)\}$ & $2\phi_K$ & $0.1718(20)\times 10^{-7}$ \\
    $8_{8,0}$ & $p_\alpha ^8$               & $F_{mmm}$ & $-0.5619(47)\times 10^{-8}$ \\
    $8_{8,0}$ & $(1-\cos 12\alpha)$          & $(1/2)V_{12}$ & $-6.9965(61)$ \\   
    $8_{7,1}$ & $p_\alpha ^7 P_a$           & $\rho_{mmm}$ & $-0.2573(24)\times 10^{-7}$ \\
    $8_{6,2}$ & $p_\alpha ^6P_a^2$          & $F_{mmK}$ & $-0.4954(50)\times 10^{-7}$ \\ 
    $8_{6,2}$ & $p^6_\alpha(P_b^2-P_c^2)$ & $F_{mmbc}$ & $0.1167(30)\times 10^{-11}$  \\
    $8_{6,2}$ & $P^2(1-\cos 9\alpha)$       & $V_{9J}$ & $0.322(11)\times 10^{-4}$ \\  
    $8_{6,2}$ & $P_a^2(1-\cos 9\alpha)$      & $V_{9K}$ & $0.22615(29)\times 10^{-3}$  \\
    $8_{6,2}$ & $(1/2)\{P_a{,}P_b\}(1-\cos 9\alpha)$ & $V_{9ab}$ & $0.15887(39)\times 10^{-3}$  \\ 
    $8_{6,2}$ & $(P_b^2-P_c^2)(1-\cos 9\alpha)$ & $V_{9bc}$ & $0.6532(89)\times 10^{-4}$ \\
    $8_{6,2}$ & $(1/2)\{P_b{,}P_c\}\sin 9\alpha$ & $D_{9bc}$ & $0.313(17)\times 10^{-4}$ \\
    $8_{5,3}$ & $p_\alpha ^5P_a^3$          & $\rho_{mmK}$ & $-0.5159(59)\times 10^{-7}$ \\  
    $8_{5,3}$ & $(1/2)\{P_a{,}P_b{,}P_c{,}p_\alpha{,} \sin 6\alpha\}$ & $\rho_{6bc}$ & $-0.2359(94)\times 10^{-6}$ \\
    $8_{4,4}$ & $p_\alpha ^4P_a^4$          & $F_{mKK}$ & $-0.3090(41)\times 10^{-7}$ \\
    $8_{4,4}$ & $(1/2)\{P_b^2{,}P_c^2\}p^4_{\alpha}$ & $F_{mb2c2}$ & $-0.1148(60)\times 10^{-12}$ \\
    $8_{4,4}$ & $P^4(1-\cos 6\alpha)$       & $V_{6JJ}$ & $-0.1827(78)\times 10^{-9}$ \\     
    $8_{4,4}$ & $P^2 P_a^2(1-\cos 6\alpha)$ & $V_{6JK}$ & $-0.20659(69)\times 10^{-7}$ \\     
    $8_{4,4}$ & $P_a^4(1-\cos 6\alpha)$     & $V_{6KK}$ & $0.3441(31)\times 10^{-7}$ \\
    $8_{4,4}$ & $P^2(P_b^2-P_c^2)(1-\cos 6\alpha)$ & $V_{6bcJ}$ & $0.8955(26)\times 10^{-9}$ \\    
    $8_{4,4}$ & $(1/2)\{P_a^2{,}(P_b^2-P_c^2)\}(1-\cos 6\alpha)$ & $V_{6bcK}$ & $0.5979(97)\times 10^{-7}$ \\    
    $8_{4,4}$ & $(1/2)P^2\{P_a{,}P_c\}\sin 6\alpha$ & $D_{6acJ}$ & $0.455(22)\times 10^{-8}$  \\  
    $8_{4,4}$ & $(1/2)\{P_a^3{,}P_c\}\sin 6\alpha$ & $D_{6acK}$ & $-0.910(23)\times 10^{-7}$  \\
    $8_{4,4}$ & $(1/2)P^2\{P_b{,}P_c\}\sin 6\alpha$ & $D_{6bcJ}$ & $0.2264(90)\times 10^{-9}$  \\ 
    $8_{4,4}$ & $(1/2)\{P_a^2{,}P_b{,}P_c{,}p_\alpha^2{,}\sin 3\alpha\}$ & $D_{3bcmK}$ & $0.300(20)\times 10^{-9}$ \\  
    $8_{3,5}$ & $p_\alpha ^3P_aP^4$         & $\rho_{mJJ}$ & $-0.1605(83)\times 10^{-13}$ \\
    $8_{3,5}$ & $p_\alpha ^3P_a^5$          & $\rho_{mKK}$ & $-0.1024(16)\times 10^{-7}$ \\       
    $8_{2,6}$ & $p^2_{\alpha} P^6$        & $F_{JJJ}$ & $0.2113(50)\times 10^{-15}$ \\
    $8_{2,6}$ & $p^2_{\alpha} P_a^6$        & $F_{KKK}$ & $-0.1522(30)\times 10^{-8}$ \\ 
    $8_{2,6}$ & $P^6(1-\cos 3\alpha)$       & $V_{3JJJ}$ & $-0.1844(29)\times 10^{-13}$ \\
    $8_{2,6}$ & $P^4P_a^2(1-\cos 3\alpha)$   & $V_{3JJK}$ & $0.1535(19)\times 10^{-11}$ \\
    $8_{2,6}$ & $P^4(P_b^2-P_c^2)(1-\cos 3\alpha)$ & $V_{3bcJJ}$ & $0.1314(11)\times 10^{-12}$ \\
    $8_{2,6}$ & $(1/2)P^2\{P_a^2{,}(P_b^2-P_c^2)\}(1-\cos 3\alpha)$ & $V_{3bcJK}$ & $-0.3552(71)\times 10^{-11}$ \\  
    $8_{2,6}$ & $(P_b^6-P_c^6)\cos 3\alpha$ & $V_{3b6c6}$ & $0.1441(11)\times 10^{-12}$  \\
    $8_{2,6}$ & $(1/2)\{P_a^5{,}P_c\}\sin 3\alpha$ & $D_{3acKK}$ & $0.555(25)\times 10^{-9}$  \\  
    $8_{2,6}$ & $(1/2)P^4\{P_b{,}P_c\}\sin 3\alpha$ & $D_{3bcJJ}$ & $0.1229(18)\times 10^{-12}$ \\  
    $8_{2,6}$ & $(1/2)\{P_b^3{,}P_c^3\}\sin 3\alpha$ & $D_{3b3c3}$ & $-0.2336(24)\times 10^{-12}$ \\ 
    $8_{2,6}$ & $(1/2)(\{P_b^5{,}P_c\}-\{P_b{,}P_c^5\})\sin 3\alpha$ & $D_{3bcbc6}$ & $0.5405(51)\times 10^{-13}$ \\ 
    $8_{1,7}$ & $(1/2)p_\alpha \{P_a^4{,}P_b\}P^2$  & $\rho_{abJK}$ & $0.272(12)\times 10^{-12}$ \\
    $8_{0,8}$ & $P^8$                       & $L_J$ & $-0.1574(73)\times 10^{-17}$ \\
    $8_{0,8}$ & $P_a^8$                       & $L_K$ & $0.1377(61)\times 10^{-10}$ \\
    $10_{8,2}$ & $P^2(1-\cos 12\alpha)$       & $V_{12J}$ & $-0.1356(27)\times 10^{-3}$ \\  
    $10_{8,2}$ & $(P_b^2-P_c^2)(1-\cos 12\alpha)$ & $V_{12bc}$ & $-0.617(27)\times 10^{-4}$ \\
    $10_{8,2}$ & $(1/2)\{P_a{,}P_c\}\sin 12\alpha$ & $D_{12ac}$ & $-0.3484(10)\times 10^{-3}$ \\
    $10_{8,2}$ & $(1/2)\{P_b{,}P_c\}\sin 12\alpha$ & $D_{12bc}$ & $-0.2484(51)\times 10^{-3}$ \\  
    $10_{7,3}$ & $(1/2)\{P_a{,}P_b{,}P_c{,}p_\alpha{,} \sin 9\alpha\}$ & $\rho_{9bc}$ & $-0.3805(87)\times 10^{-5}$ \\
    $10_{6,4}$ & $P^4(1-\cos 9\alpha)$       & $V_{9JJ}$ & $0.1013(14)\times 10^{-8}$ \\  
    $10_{6,4}$ & $(1/2)\{P_a^2{,}P_b{,}P_c\}\sin 9\alpha$ & $D_{9bcK}$ & $-0.3058(58)\times 10^{-5}$  \\ 
    $10_{4,6}$ & $P^6(1-\cos 6\alpha)$       & $V_{6JJJ}$ & $-0.546(34)\times 10^{-14}$ \\
    $10_{4,6}$ & $P^4(P_b^2-P_c^2)(1-\cos 6\alpha)$ & $V_{6bcJJ}$ & $-0.1489(60)\times 10^{-13}$ \\
    $10_{2,8}$ & $(1/2)P^6\{P_a{,}P_c\}\sin 3\alpha$ & $D_{3acJJJ}$ & $0.206(13)\times 10^{-15}$  \\  
    $10_{2,8}$ & $(1/2)\{P_a^6{,}P_b{,}P_c\}\sin 3\alpha$ & $D_{3bcKKK}$ & $-0.1237(38)\times 10^{-11}$  \\
    $12_{8,4}$ & $(1/2)\{P_a^3{,}P_c\}\sin 12\alpha$ & $D_{12acK}$ & $0.2173(58)\times 10^{-6}$  \\  
    $12_{8,4}$ & $(1/2)\{P_a^2{,}P_b{,}P_c\}\sin 12\alpha$ & $D_{12bcK}$ & $0.1723(47)\times 10^{-5}$  \\  
    $12_{4,8}$ & $(1/2)\{P_a^6{,}P_b{,}P_c\}\sin 6\alpha$ & $D_{6bcKKK}$ & $0.587(26)\times 10^{-12}$  \\  

\hline
\end{longtable}
\tablefoot{$^{a}$ $n=t+r$, where $n$ is the total order of the operator, $t$ is the order of the torsional part 
and $r$ is the order of the rotational part, respectively. The ordering scheme of \citet{NAKAGAWA:1987} is used. 
$^{b}$ \{A,B,C,D,E\} = ABCDE + EDCBA. \{A,B,C,D\} = ABCD + DCBA. \{A,B,C\} = ABC + CBA. \{A,B\} = AB + BA. The product of the operator 
in the second column of a given row and the parameter in the third column of that row gives the term actually used 
in the torsion-rotation Hamiltonian of the program, except for $F$, $\rho$ and $A_{\rm RAM}$, 
which occur in the Hamiltonian in the form $F(p_a + \rho P_a)^2 + A_{\rm RAM}P_a^2$. $^{c}$ The parameter nomenclature 
is based on the subscript procedure of \citet{XU:2008305}. $^{d}$ Values of the parameters in cm$^{-1}$, 
except for $\rho$, which is unitless. $^{e}$ Statistical uncertainties are given in parentheses as one standard 
uncertainty in units of the last digits.}
}

\end{appendix}

\end{document}